\documentclass{PoS}

\usepackage{cite}

\newcommand{\be}{\begin{equation}}
\newcommand{\ee}{\end{equation}}
\newcommand{\bea}{\begin{eqnarray}}
\newcommand{\eea}{\end{eqnarray}}

\newcommand{\bmp}[1]{\begin{minipage}{#1cm}}
\newcommand{\emp}{\end{minipage}}

\newcommand{\bra}{\langle}
\newcommand{\ket}{\rangle}

\newcommand{\bean}{\begin{eqnarray*}}
\newcommand{\eean}{\end{eqnarray*}}

\title{Spectral quantities in thermal QCD: a progress report from the FASTSUM collaboration}

\ShortTitle{Spectral quantities in thermal QCD}

\author{\speaker{Gert Aarts}, Chris Allton, Jonas Glesaaen, Simon Hands, Aleksandr Nikolaev \\
    Department of Physics, Swansea University, Swansea SA2 8PP, United Kingdom\\
     E-mail: 
     \email{\{g.aarts,c.allton,s.hands,aleksandr.nikolaev\}@swan.ac.uk, \\jonas@glesaaen.com}}

\author{Benjamin J\"ager\\
        CP3-Origins \& Danish IAS, Department of Mathematics and
	Computer Science, University of Southern Denmark, 5230 Odense M, Denmark \\
	E-mail: 
	\email{jaeger@cp3.sdu.dk}}

\author{Seyong Kim\\
        Department of Physics, Sejong University, Seoul 143-747, Korea\\
        E-mail: 
        \email{skim@sejong.ac.kr}}

\author{Maria Paola Lombardo\\
        INFN, Sezione di Firenze, 50019 Sesto Fiorentino (FI), Italy\\
        E-mail: 
        \email{Mariapaola.Lombardo@lnf.infn.it}}

\author{Sinead M.\ Ryan\\
	School of Mathematics, Trinity College, Dublin 2, Ireland\\
	E-mail: 
	\email{ryan@maths.tcd.ie}}

\author{Jonivar Skullerud\\
        Dept.\ of Theoretical Physics, National University of Ireland Maynooth, County Kildare, Ireland\\
        E-mail: 
        \email{jonivar@thphys.nuim.ie}}

\author{Liang-Kai Wu\\
        Faculty of Science, Jiangsu University, Zhenjiang, 212013 \& Key Laboratory of Quark and Lepton Physics (MOE), Central China Normal University,
Wuhan 430079, China\\
        E-mail: 
        \email{wuliangkai@ujs.edu.cn}}

\abstract{
In order to study spectral quantities in thermal QCD, the FASTSUM collaboration employs anisotropic lattice simulations with $N_f=2+1$ flavours of Wilson fermions. Here we discuss our Generation 2 and Generation 2L ensembles, which differ in the pion mass. The focus is on observables related to the light quarks and chiral symmetry restoration.
}

\FullConference{37th International Symposium on Lattice Field Theory - Lattice2019\\
		16-22 June 2019\\
		Wuhan, China}

\begin{document}

\section{Introduction}

The study of strongly interacting matter under extreme conditions remains a topical challenge in thermal lattice QCD. The aim of the FASTSUM collaboration  \cite{fastsum} is to go beyond bulk thermodynamic quantities and arrive at a detailed understanding of spectral quantities in both the hadronic and quark-gluon plasma phases. By providing a thorough analysis of mesonic and baryonic channels over a range of temperatures, a comprehensive understanding can be obtained, linked with e.g.\  manifestations of chiral symmetry restoration for light quarks, interquark potentials for heavy ones, and transport for conserved currents. Such a study is relevant for the phenomenology of heavy-ion collisions, where (possibly flavour-dependent) in-medium effects play an important role.

From the perspective of simulations, these studies are still at a more exploratory stage than is the case for QCD bulk thermodynamics. While for the latter \cite{Borsanyi:2010cj,Bazavov:2014pvz}, simulations are carried out at the physical point and for a sequence of lattice spacings, permitting an extrapolation to the continuum limit, in our programme we are currently working away from the physical point and at a single spatial lattice spacing. There are (at least) two reasons for this, both related to having access to a sufficient number of lattice points in the temporal direction: we employ Wilson-type quarks rather than staggered ones, often used for thermodynamics, to ensure every time slice contributes in the analysis; we use anisotropic lattices, with $a_\tau<a_s$, to increase temporal resolution. The latter implies that a new tuning of the anisotropy parameters is required each time the lattice spacing is changed.
 In this contribution, we update the status of our project, moving to lighter quarks, while staying at approximately the same lattice spacings, with $a_s/a_\tau\sim 3.5$. Compared to last year \cite{Aarts:2018haw}, we have increased the statistics and made progress on the analysis of the ensembles, especially at the lower temperatures. Here we report on our findings for susceptibilities, the chiral condensate, baryons and parity doubling. In two related contributions,  ${\cal O}(\mu^2)$ corrections to the light meson spectrum \cite{Aleksandr} and bottomonium on the new ensembles \cite{Sam} are discussed. Other work considering QCD thermodynamics with Wilson-type quarks includes Refs.\ \cite{Borsanyi:2012uq,Taniguchi:2016ofw,Burger:2018fvb,Kanaya:2019okb}.

\section{FASTSUM ensembles}

We consider $N_f=2+1$ dynamical quark flavours, of the Wilson-clover type, on anisotropic lattices with  $\xi=a_s/a_\tau \approx 3.5$. The tuning of the anisotropy and the ensembles at the lowest temperature (``$T=0$'') are made available through the HadSpec collaboration \cite{Edwards:2008ja,Wilson:2015dqa}  and denoted with a $^*$ below. We work at a single fixed lattice spacing. While the strange quark mass $m_s$ is at its physical value, the light quarks are not yet at the physical point.  En route from $m_q\to m_{ud}$, we consider here our {\em Generation} 2L ensembles, with $m_\pi = 236(2)$ MeV, and contrast them with our previous {\em Generation} 2 ensembles, with $m_\pi = 384(4)$ MeV  \cite{Aarts:2014cda,Aarts:2014nba}. We employ \cite{openqcd-fastsum} a modification of OpenQCD \cite{openqcd}, to include anisotropy and stout-smearing, supplemented with a stand-alone spectroscopy code \cite{openqcd-hadspec}.
Details of the Gen 2L ensembles are given in Table~\ref{tab2L}, while those for Gen 2 can be found in Refs.\ \cite{Aarts:2018haw,Aarts:2014cda,Aarts:2014nba}.

\begin{table}[t]
\begin{center}
\begin{tabular}{|c|| ccccccc |}
\hline
$N_\tau$        & 256$^*$& 128  & 64    & 56    & 48    & 40    & 36 \\ \hline
$T$ [MeV]  		& 23    & 47    & 94    & 107   & 125   & 150   & 167 \\
$N_{\rm cfg}$ 	& 750   & 306   & 1041  & 1042  & 1123  & 1102  & 1119 \\
\hline
$N_\tau$        & 32    & 28    & 24    & 20    & 16    & 12    & 8 \\    \hline
$T$ [MeV]  		& 187   & 214   & 250   & 300   & 375   & 500   & 750\\
$N_{\rm cfg}$ 	& 1090  & 1031  & 1016  & 1030  & 1102  & 1267  & 1048  \\
\hline
\end{tabular}
\caption{Gen 2L ensembles, $m_\pi=236(2)$ MeV, lattice size $32^3 \times N_\tau$, spatial lattice spacing $a_s=0.1136(6)$ fm, temporal lattice spacing $a_\tau^{-1}=5.997(34)$ GeV, anisotropy $a_s/a_\tau=3.453(6)$.  }
\label{tab2L}
\end{center}
\end{table}

\section{Fluctuations and chiral properties}

In order to interpret the results from the spectral studies at finite temperature, it is necessary to first understand properties of the thermal crossover for Wilson fermions. 
Here we give some results linked to fluctuations and chiral properties, comparing the Gen 2 and 2L ensembles. Since the dominant difference is a reduction of the pion mass, it is expected that the pseudocritical temperatures, marking the crossover, shift to lower values as $m_\pi$ is reduced.

\begin{figure}[b]
\begin{center}
  \includegraphics[width=0.49\textwidth]{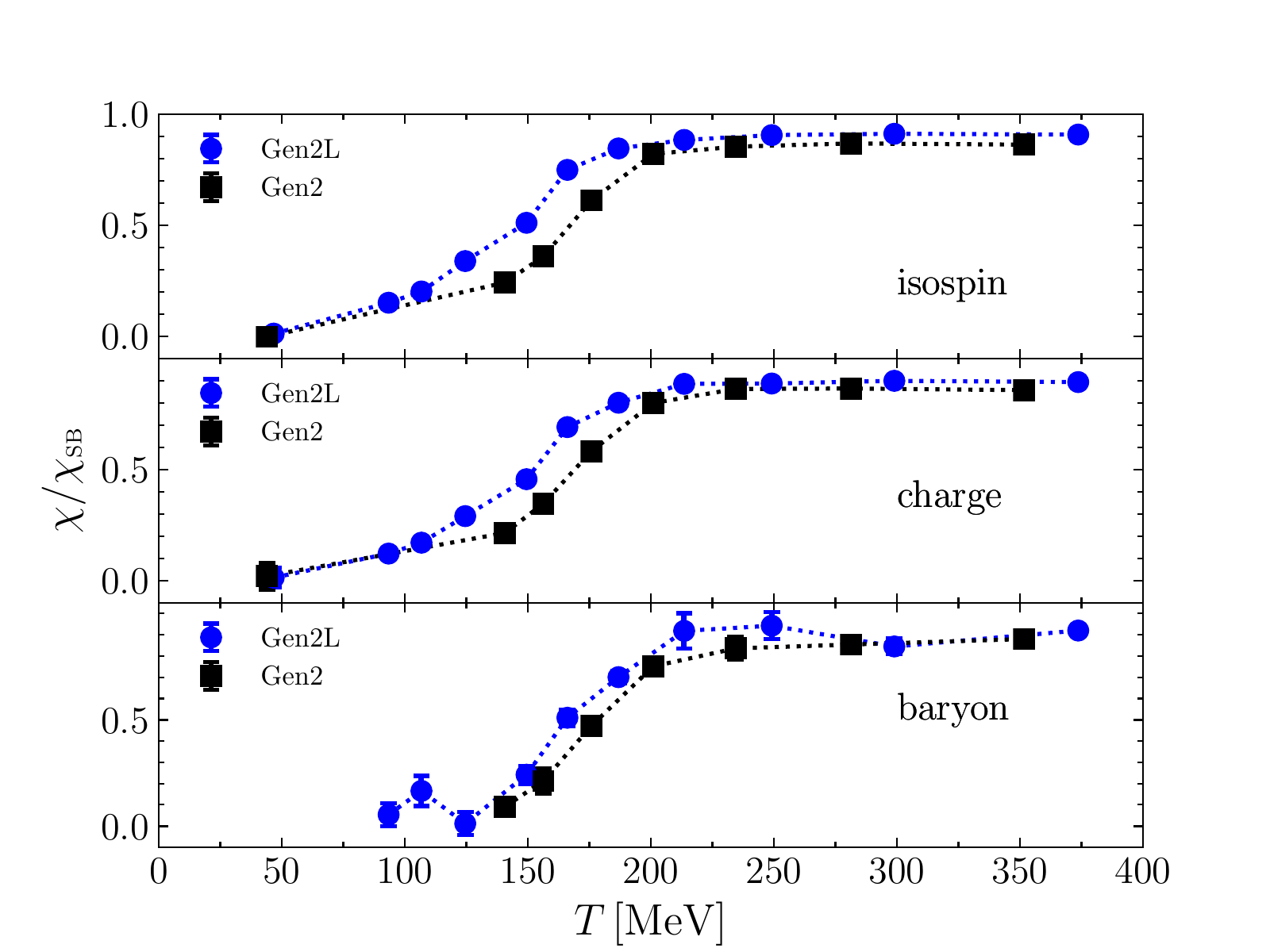}
  \includegraphics[width=0.49\textwidth]{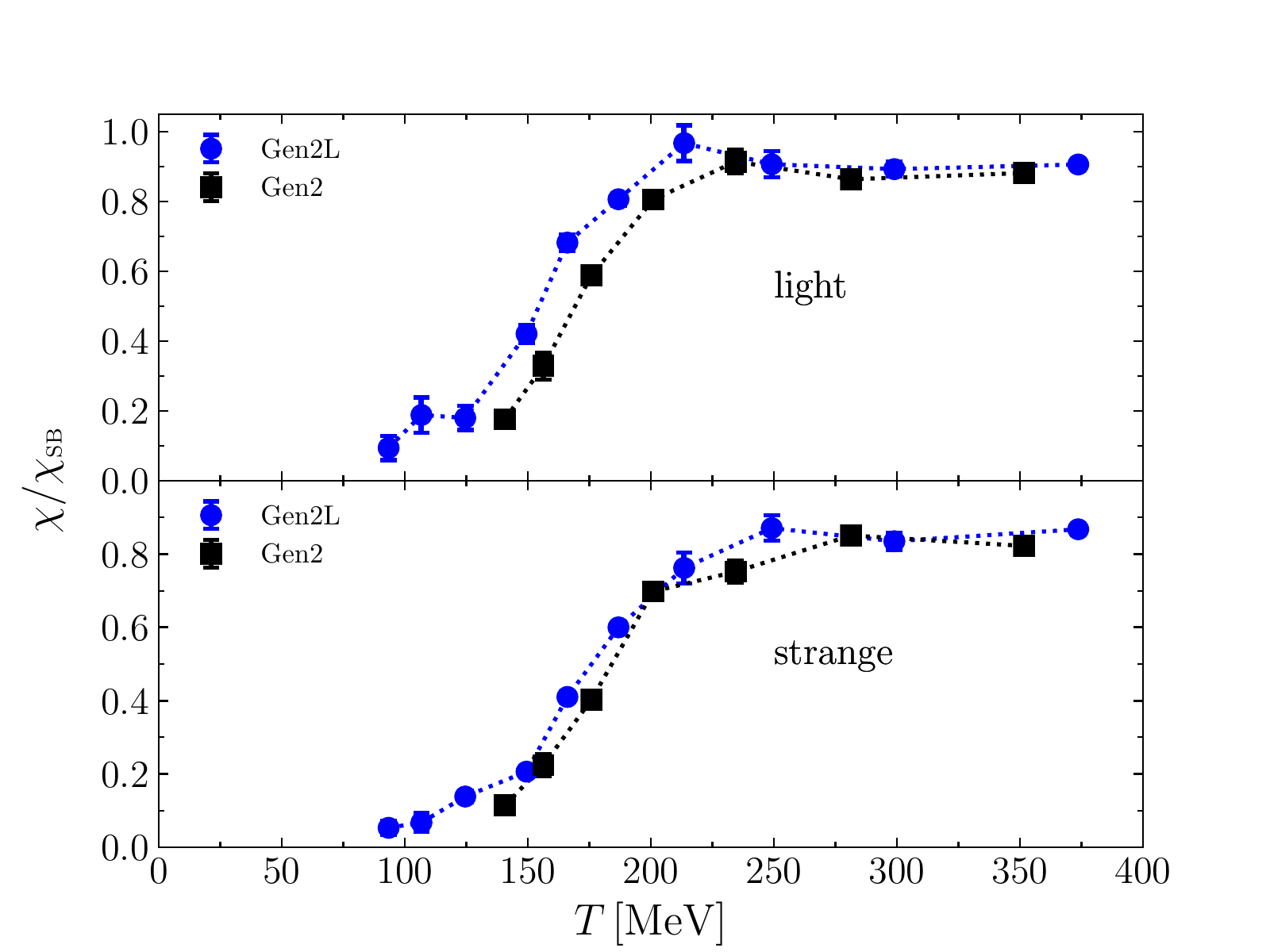}
\caption{Susceptibilities $\chi/\chi_{\rm SB}$ for isospin (I), electrical charge (Q) and baryon number (B) (left), and light and strange quark number (right), normalised with the massless, lattice Stefan-Boltzmann result, for Gen 2 and 2L. The lines are meant to guide the eye.
\label{fig:sus}
}
\end{center}
\end{figure}

In Fig.\ \ref{fig:sus} we show the susceptibilities representing fluctuations of isospin, electric charge and baryon number, and light and strange quark number, normalised with the lattice Stefan-Boltzmann result for massless quarks, comparing Gen 2 \cite{Aarts:2014nba} and 2L. The most important result is the shift of the transition to lower temperatures, as expected. This can be quantified by fitting the data with cubic splines and determining the temperature of the inflection point, to define pseudocritical temperatures. The findings are summarised in Table \ref{tab:Tc}.

\begin{figure}[t]
\begin{center}
 \includegraphics[width=0.49\textwidth]{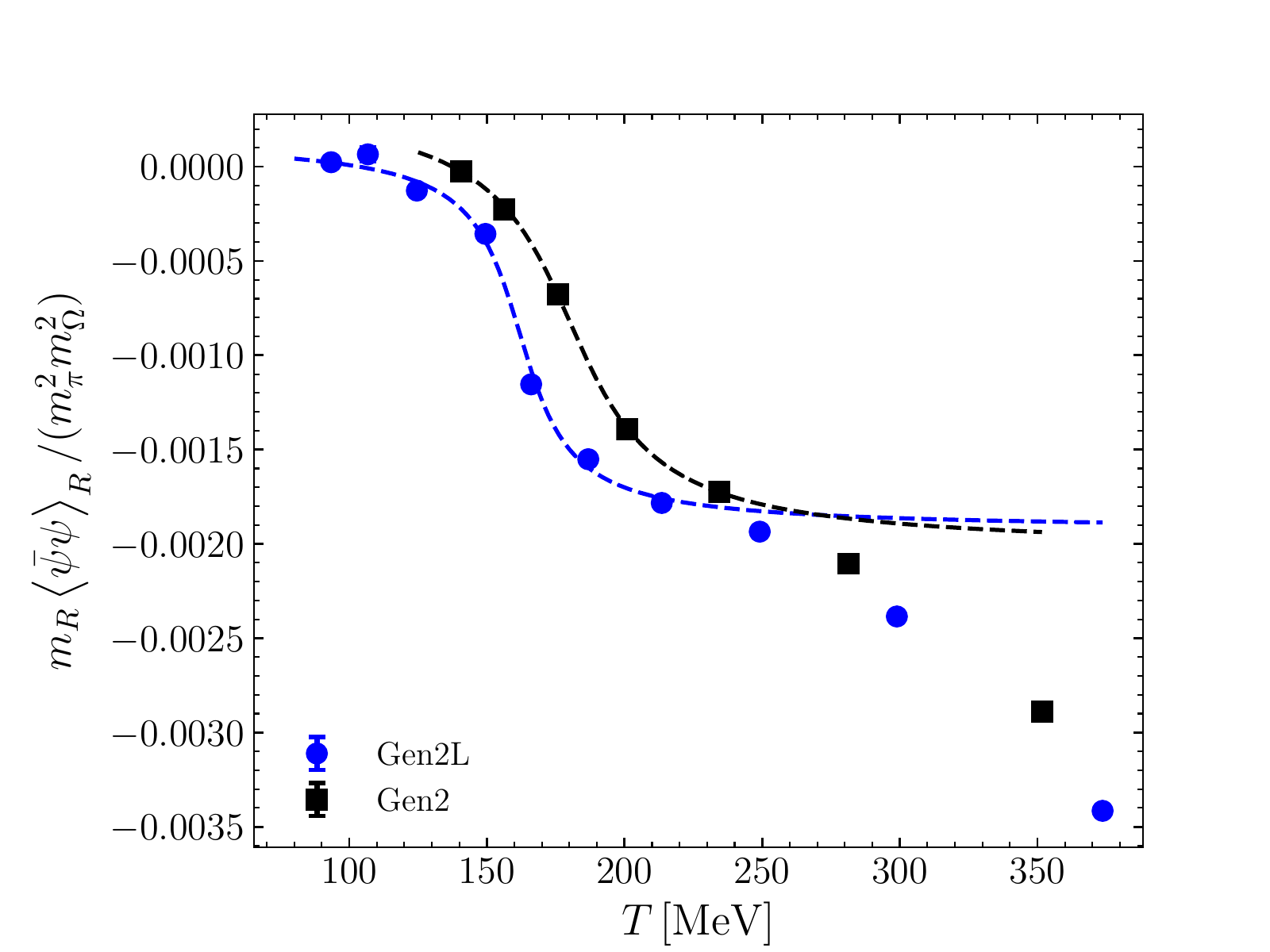}
 \includegraphics[width=0.49\textwidth]{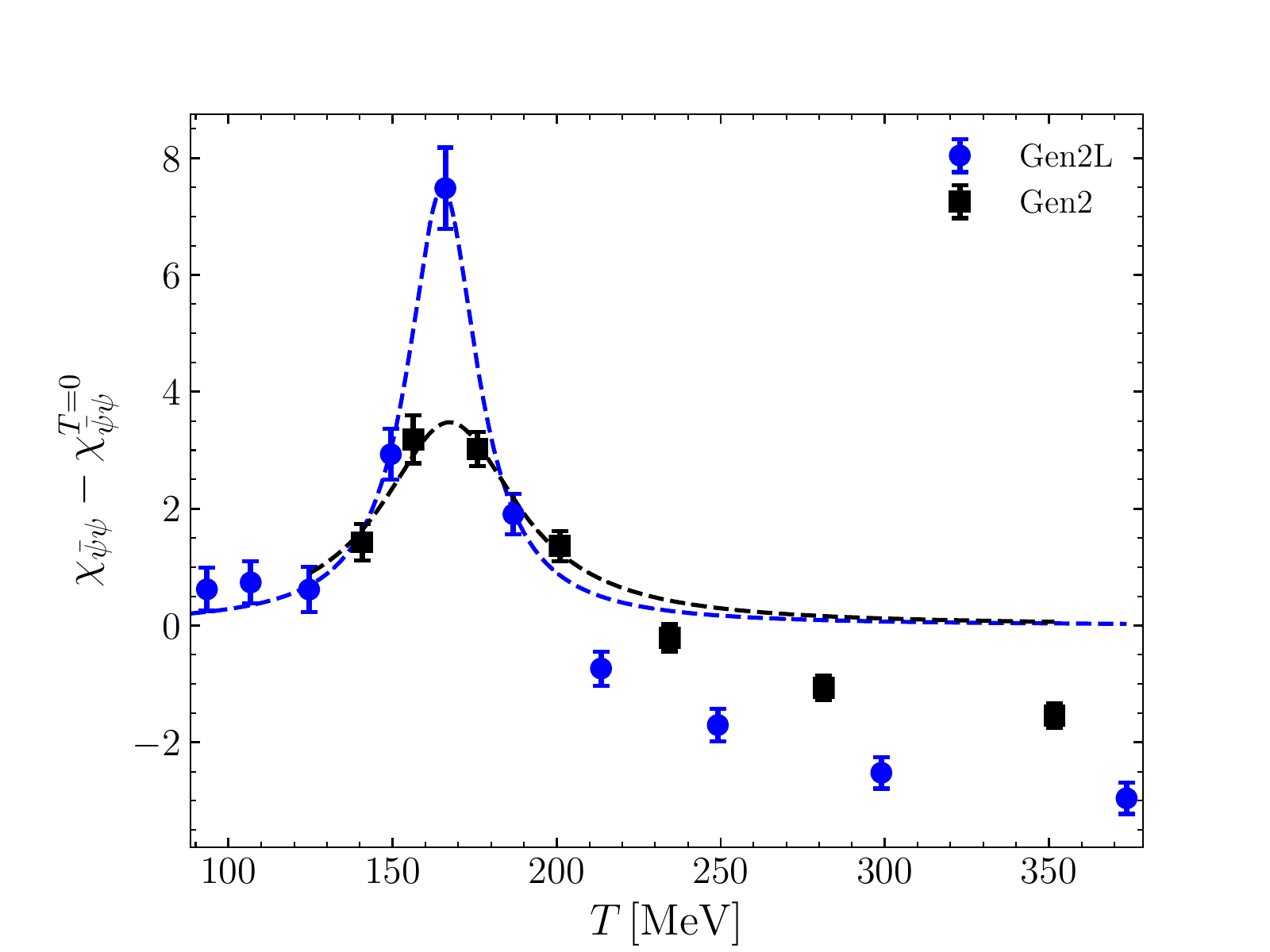}
  \caption{Renormalised chiral condensate (left) and subtracted chiral susceptibility (right), for Gen 2 and~2L.
  \label{fig:cc}}
\end{center}
\end{figure}

In addition, we have studied the chiral condensate and susceptibility for light quarks, defined via the first respectively second derivative of the free energy with respect to the quark mass.
In order to renormalise these, we note that we use a fixed-scale approach, in which additive and multiplicative renormalisation factors are independent of the temperature. We first discuss the susceptibility, see Fig.\ \ref{fig:cc} (right). To remove the additive divergence, we subtract the value at the lowest temperature. The remaining multiplicative factor is the same at all temperatures, but not the same for Gen 2 and 2L, which  may explain the different behaviour at large temperatures; it can indeed be absorbed by a simple rescaling. The interest is in the peak position, which is extracted using a simple fit according to $A/[B^2+(T-T_{\rm pc})^2]$. The result for this pseudocritical temperature is given in Table \ref{tab:Tc} as well, labelled by $\chi_{\bar\psi\psi}$. We note that for Gen 2 the peak is broader and less pronounced,
as expected of a pseudocritical scaling, making the estimate less reliable.

\begin{table}[t]
\begin{center}
    \begin{tabular}{| c|| c | c || c||  c | c |}
    \hline
                            & \multicolumn{2}{c|| }{$T_{\rm pc}$ [MeV]} & & \multicolumn{2}{c|}{$T_{\rm pc}$ [MeV]}  \\
         \hline
         observable          & Gen 2    & Gen 2L    & observable            & Gen 2     & Gen 2L \\        \hline   
        $\chi_{\rm I}$       & 168.0(2) & 157.2(1)  & $\chi_{\rm light}$    & 162(2)    & 154(1)  \\
        $\chi_{\rm Q}$       & 167.9(8) & 157.3(3)  & $\chi_{\rm strange}$  & 184(1)    & 161(1)  \\
        $\chi_{\rm B}$       & 172(2)   & 153(1)    & $\bra\bar\psi\psi\ket_R$& 181(1)  & 162(1) \\
        $\chi_{\bar\psi\psi}$& 167(3)   & 165(2)    & $R$                   & 164-177   & 159-161  \\
        \hline
    \end{tabular}
    \caption{Pseudocritical temperatures extracted via various susceptibilities $\chi$, the renormalised chiral condensate $\bra\bar\psi\psi\ket_R$, and the parity doubling
parameter $R$ for octet and decuplet baryons. The latter depends on strangeness in Gen 2, but significantly less so in Gen 2L.
    \label{tab:Tc}}
\end{center}
\end{table}

For the chiral condensate, we follow the procedure of Refs. \cite{Borsanyi:2012uq, Giusti:1998wy}, which yields a renormalized result for $m_R\bra\bar\psi\psi\ket_R$, where $m_R$ denotes the renormalised quark mass. To make this quantity dimensionless and well-defined in the chiral limit, we normalise it with $m_\pi^2 m_\Omega^2$, using the ``$T=0$'' values for Gen2 and 2L respectively. The result is shown in Fig.\ \ref{fig:cc} (left). We observe a shift of the inflection point towards lower temperatures. Following Ref.\ \cite{Burger:2018fvb}, we fit the temperature dependence according to $A+B\arctan[C(T-T_{\rm pc})]$, including only the data points in the $S$-shaped part, which yields the values for the pseudocritical temperature given in Table~\ref{tab:Tc}. The observed drop of the value of the condensate at higher temperatures is currently under investigation. 

\begin{figure}[t]
  \begin{center}
  \includegraphics[width=0.5\textwidth]{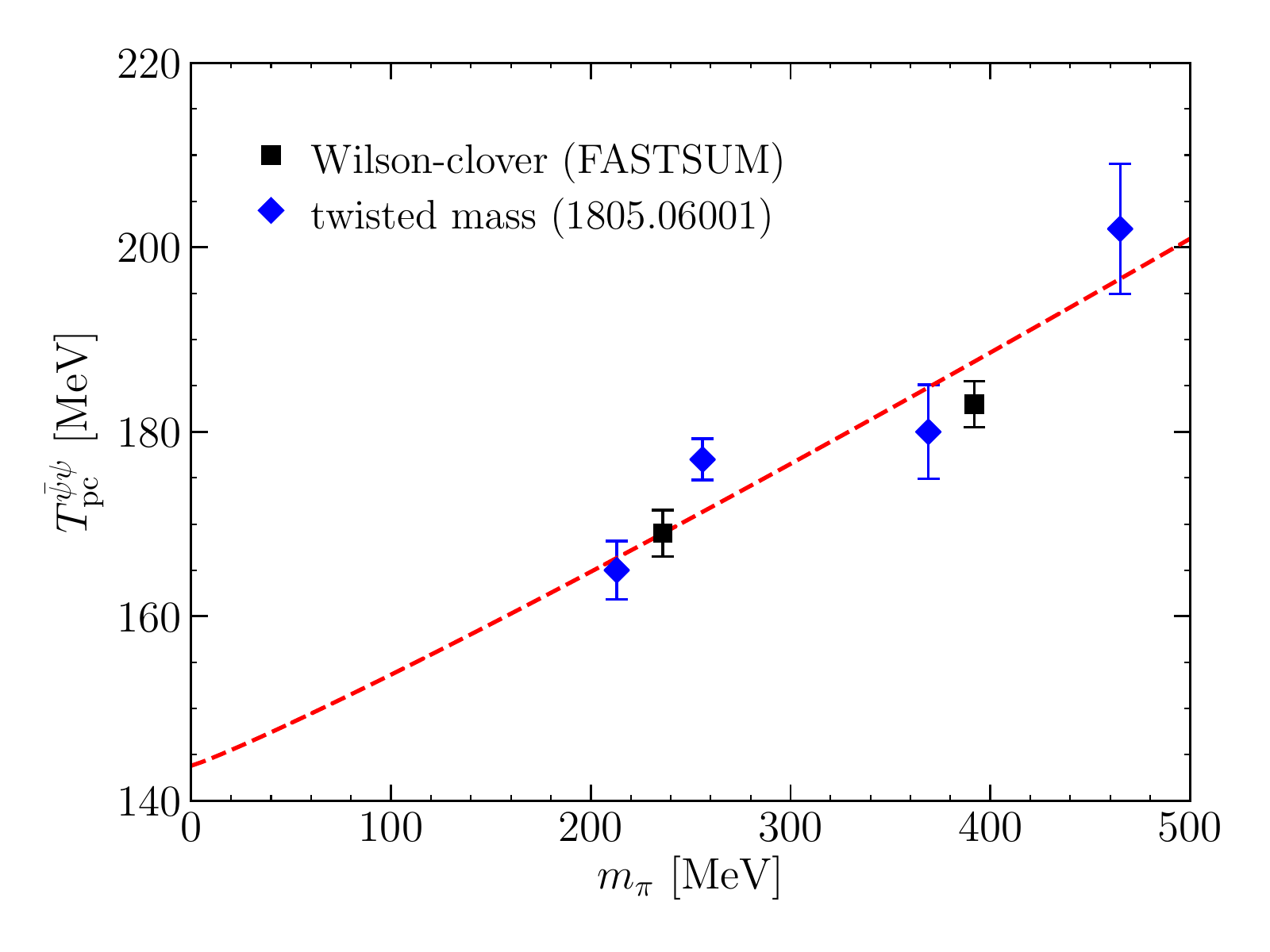}
  \vspace*{-0.3cm}
  \caption{Estimates for the pion mass dependence of $T_{\rm pc}$, extracted from the inflection point of the renormalised chiral condensate, for twisted-mass \cite{Burger:2018fvb} and Wilson-clover (this work) fermions.
  \label{fig:Tc}}
  \end{center}

\vspace*{-1cm}

  \begin{center}
  \includegraphics[width=0.49\textwidth]{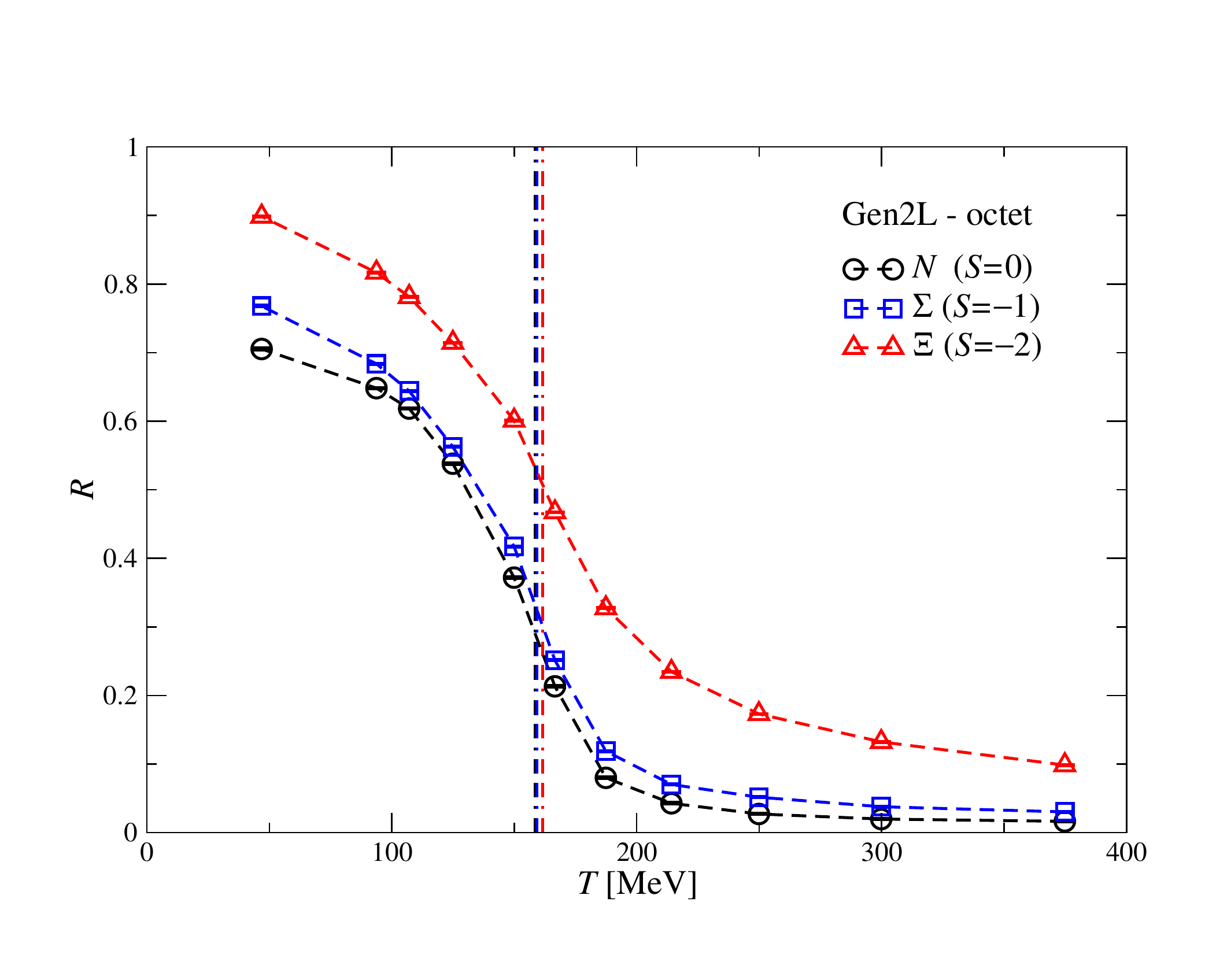}
  \includegraphics[width=0.49\textwidth]{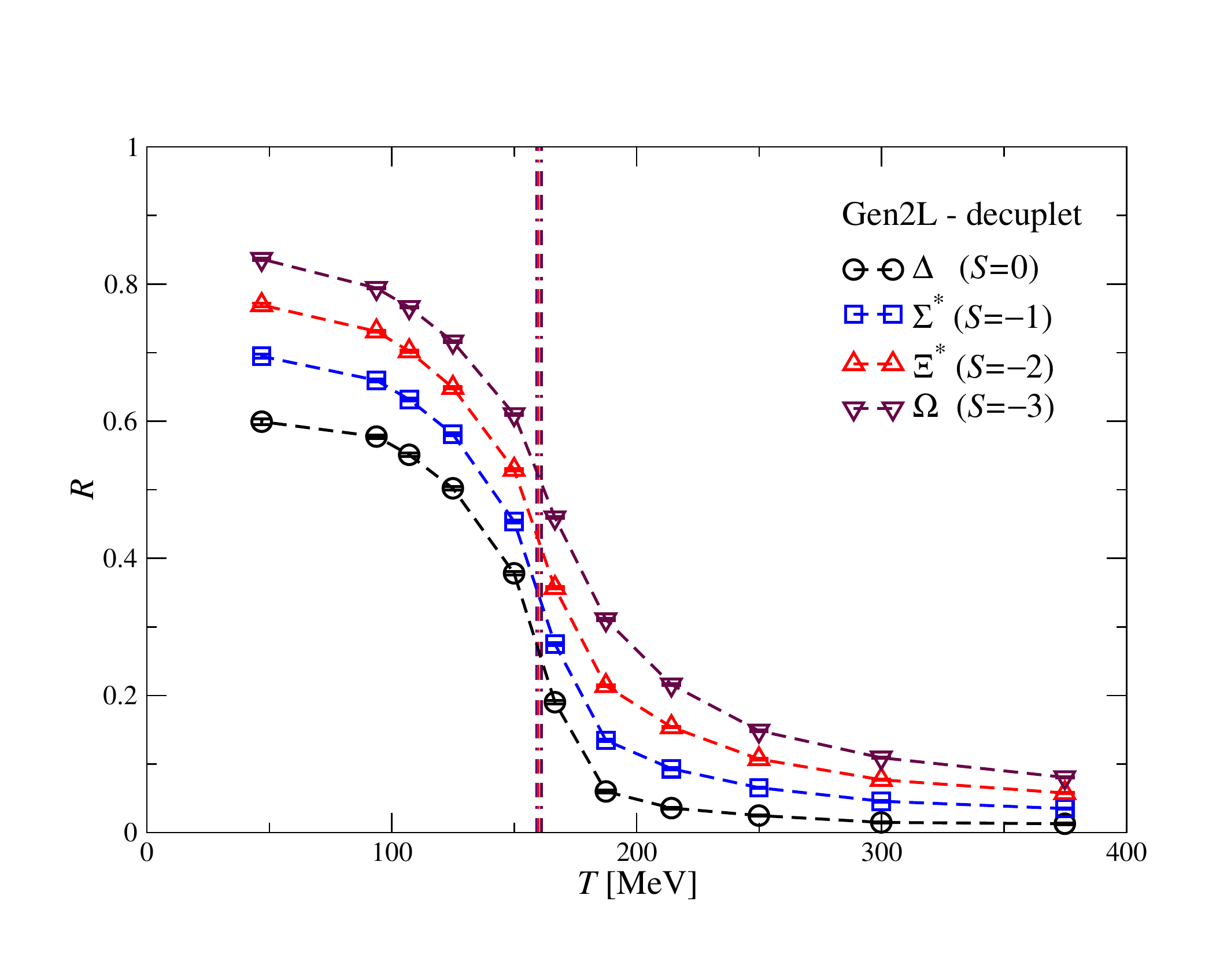}
  \vspace*{-0.3cm}
  \caption{Parity doubling
parameter $R$ for octet (left) and decuplet (right) baryons as a function of temperature. The vertical lines indicate the inflection points.
  \label{fig:R}}
    \end{center}
\end{figure}

Unlike in the case of staggered quarks, thermal simulations with Wilson-type fermions do not yet take place at the physical point. Hence it is important to compare the pion mass dependence of our results with those in the literature. This is presented in Fig.\ \ref{fig:Tc} for the renormalised chiral condensate, where results from twisted-mass fermions \cite{Burger:2018fvb} are compared to ours. We note a good consistency between the different Wilson-type formulations, providing support for our findings.


\begin{figure}[t]
  \begin{center}
  \includegraphics[width=0.49\textwidth]{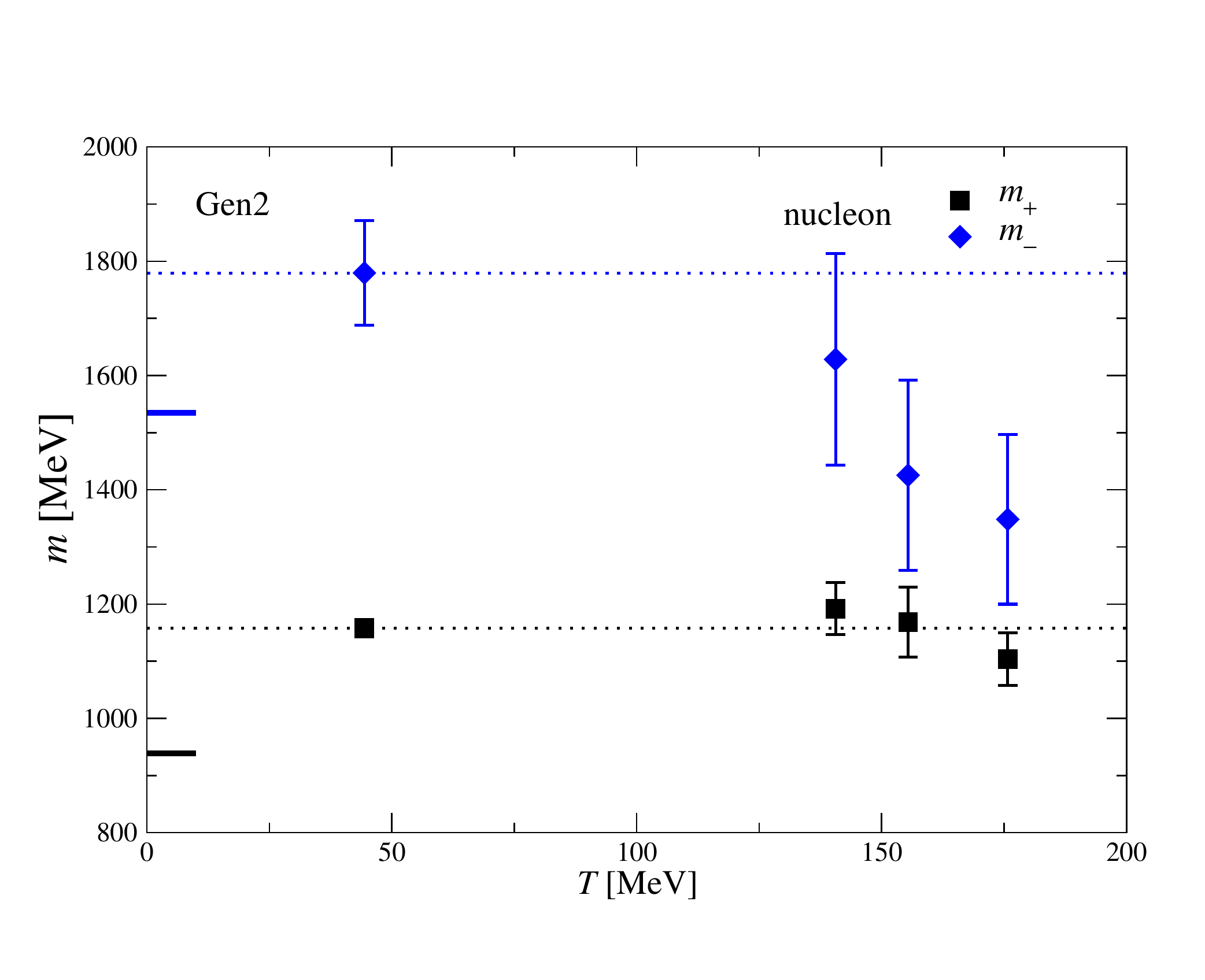}
  \includegraphics[width=0.49\textwidth]{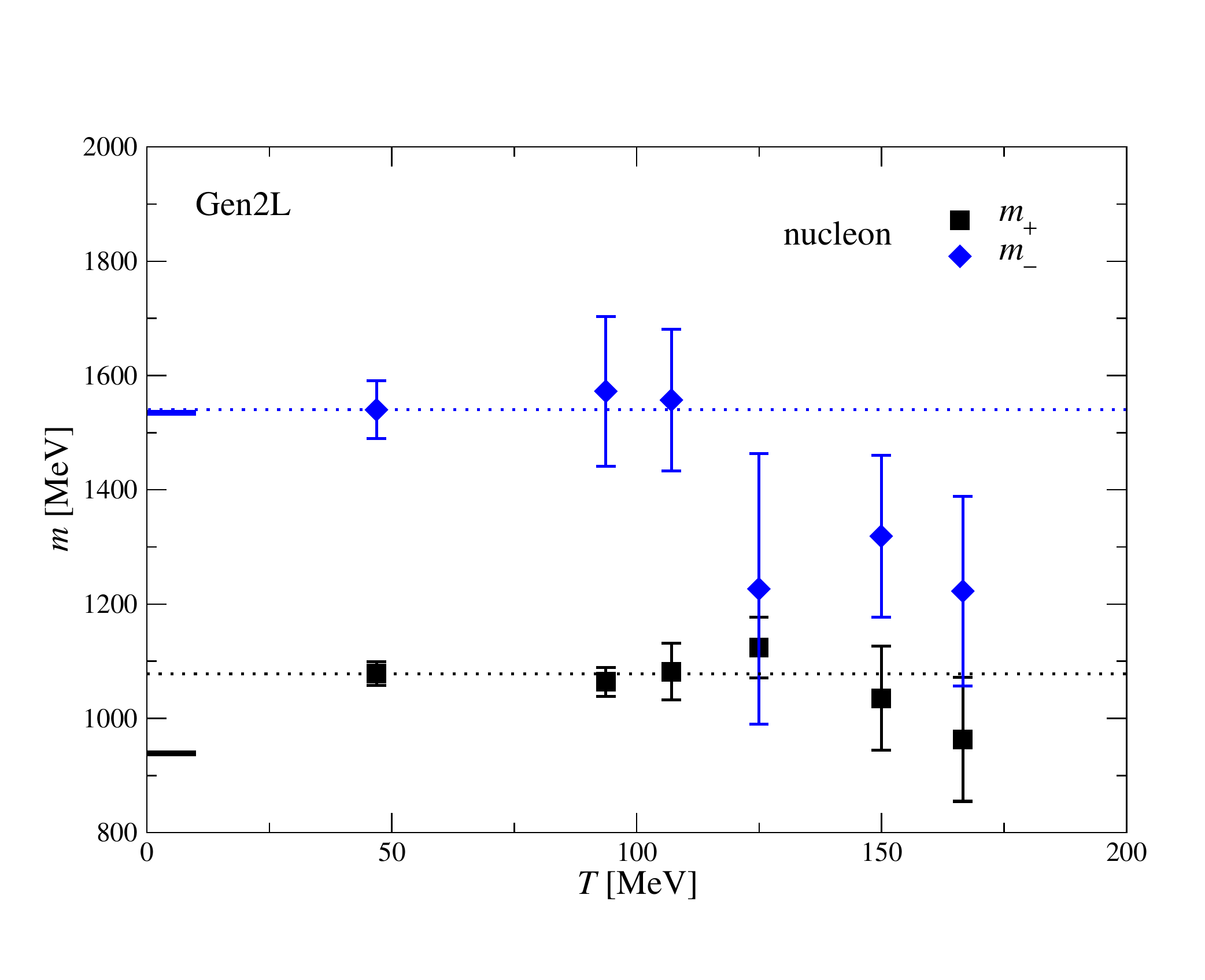}
  \vspace*{-0.3cm}
  \caption{Temperature dependence of the nucleon ground state masses $m_\pm$  in the two parity channels, for Gen 2 (left) and 2L (right). Dotted lines indicate the value at the lowest temperature and are meant to guide the eye, while the stubs sticking out on the left are the physical values.
  \label{fig:mass}}
    \end{center}
\end{figure}

\section{Baryons and parity doubling}

As a second topic, we consider parity doubling in the light baryon sector, previously discussed for the Gen 2 ensembles in Refs.\ \cite{Aarts:2015mma,Aarts:2017rrl,Aarts:2018glk}, and its relation to chiral symmetry breaking/restoration.
Starting from the euclidean correlator $G_\pm(\tau)$ for positive- and negative-parity states, we construct a quasi-order parameter $R$ from the ratio $R(\tau) = [G_+(\tau)-G_-(\tau)]/[G_+(\tau) + G_-(\tau)]$, see Refs.~\cite{Datta:2012fz,Aarts:2015mma,Aarts:2017rrl,Aarts:2018glk} for details.
In the case of parity doubling, i.e.\ no chiral symmetry breaking, this ratio tends to 0, while it is positive when the respective ground states satisfy $m_-\gg m_+$. The results for Gen 2L are shown in Fig.\ \ref{fig:R}. As before \cite{Aarts:2018glk}, we observe a systematic drop as the temperature increases and an ordering at high temperature according to strangeness, due to
the nonzero strange quark mass. However, while for Gen 2 the inflection points, indicating the transition, show some dependence on the strangeness content \cite{Aarts:2018glk}, here we find that the inflection points essentially coincide, at the lower temperature interval 159-161 MeV, as indicated in Table \ref{tab:Tc}. 
Overall we find therefore that the pseudocritical temperature from all observables shifts to lower values as the pion mass is reduced and, moreover, that the strangeness dependence is reduced.

As a final result, we present the temperature dependence of the nucleon ground state masses in the hadronic phase, comparing the results for Gen 2 \cite{Aarts:2018glk} with those in Gen 2L, see Fig.\ \ref{fig:mass}. Due to the light quarks being closer to the physical value, we observe a better agreement with nature at the lowest temperature, indicated with the short stubs on the left. In Gen 2L we are able to note the absence of temperature dependence up to at least $T=100$ MeV. At higher temperatures the emergence of parity doubling can be observed, following the trend found in Ref.\ \cite{Aarts:2018glk}. Interestingly, a more pronounced drop close to the transition to the deconfined phase is visible also for $m_+$, albeit with large uncertainty. A full analysis of all light baryons will be given elsewhere.

\section{Summary}

An overview was presented of FASTSUM's anisotropic $N_f=2+1$ ensembles, comparing Generations 2 and 2L, with $m_\pi = 384(4)$ and $236(2)$ MeV respectively, with a focus on properties of the chiral crossover with Wilson fermions. A full study of spectroscopy is currently in progress.

 
 \vspace*{0.5cm}
 
\noindent
{\bf Acknowledgments.}
  We are grateful for support from STFC via grants ST/L000369/1 and ST/P00\-055X/1, the Swansea Academy for Advanced Computing, SNF, ICHEC, COST Action CA15213 THOR, the European Research Council (ERC) under the European Union's Horizon 2020 research and innovation programme under grant agreement No 813942, and the Key Laboratory of Ministry of Education of China under Grant No.\ QLPL2018P01.
  Computing resources were made available by HPC Wales, Supercomputing Wales, and PRACE via access to the Marconi-KNL system hosted by CINECA, Italy.
This work used the DiRAC Extreme Scaling service and the DiRAC Blue Gene Q Shared Petaflop system at the University of Edinburgh, operated by the Edinburgh Parallel Computing Centre on behalf of the STFC DiRAC HPC Facility. This equipment was funded by by BIS National E-infrastructure capital grant ST/K000411/1, 
STFC capital grants ST/H008845/1 and ST/R00238X/1, 
and
STFC DiRAC Operations grants ST/K005804/1, ST/K005790/1 and ST/R001006/1. 
DiRAC is part of the National e-Infrastructure. 



\begin{thebibliography}{99}


\bibitem{fastsum}
    FASTSUM collaboration, \href{http://fastsum.gitlab.io/}{\tt fastsum.gitlab.io/}
   
   

\bibitem{Borsanyi:2010cj}
  S.~Bors\'anyi {\em et al.},
  JHEP {\bf 1011} (2010) 077
  [arXiv:1007.2580 [hep-lat]].

\bibitem{Bazavov:2014pvz}
  A.~Bazavov {\it et al.} [HotQCD Collaboration],
  Phys.\ Rev.\ D {\bf 90} (2014) 094503
  [arXiv:1407.6387 [hep-lat]].
  
  
  
\bibitem{Aarts:2018haw}
  G.~Aarts, C.~Allton, J.~Glesaaen, S.~Hands, B.~J\"ager and J.~Skullerud,
  PoS LATTICE {\bf 2018} (2018) 183
  [arXiv:1812.08151 [hep-lat]].

\bibitem{Aleksandr}
 A.~Nikolaev {\em et al.}, 
 PoS LATTICE {\bf 2019} (2019) 077.
 
\bibitem{Sam}
 S.~Offler {\em et al.},
 PoS LATTICE {\bf 2019} (2019) 076.
 
 

\bibitem{Borsanyi:2012uq}
  S.~Bors\'anyi {\it et al.},
  JHEP {\bf 1208} (2012) 126
  [arXiv:1205.0440 [hep-lat]];
  Phys.\ Rev.\ D {\bf 92} (2015) no.1,  014505
  [arXiv:1504.03676 [hep-lat]].

\bibitem{Taniguchi:2016ofw}
  Y.~Taniguchi {\em et al.},
  Phys.\ Rev.\ D {\bf 96} (2017) no.1,  014509;
   Erratum: [Phys.\ Rev.\ D {\bf 99} (2019) no.5,  059904]
  [arXiv:1609.01417 [hep-lat]].
  
\bibitem{Burger:2018fvb}
  F.~Burger, E.~M.~Ilgenfritz, M.~P.~Lombardo and A.~Trunin,
  Phys.\ Rev.\ D {\bf 98} (2018) no.9,  094501
  [arXiv:1805.06001 [hep-lat]].

\bibitem{Kanaya:2019okb}
  K.~Kanaya, A.~Baba, A.~Suzuki, S.~Ejiri, M.~Kitazawa, H.~Suzuki, Y.~Taniguchi and T.~Umeda,
  PoS LATTICE {\bf 2019} (2019) 088
  [arXiv:1910.13036 [hep-lat]].
  
  
  
 

\bibitem{Edwards:2008ja}
  R.~G.~Edwards, B.~Joo and H.~W.~Lin,
  Phys.\ Rev.\ D {\bf 78} (2008) 054501
  [arXiv:0803.3960 [hep-lat]];
  H.~W.~Lin {\it et al.} [Hadron Spectrum Collaboration],
  Phys.\ Rev.\ D {\bf 79} (2009) 034502
  [arXiv:0810.3588 [hep-lat]].


\bibitem{Wilson:2015dqa}
  D.~J.~Wilson, R.~A.~Bri\~ceno, J.~J.~Dudek, R.~G.~Edwards and C.~E.~Thomas,
  Phys.\ Rev.\ D {\bf 92} (2015) no.9,  094502
  [arXiv:1507.02599 [hep-ph]];
  G.~K.~C.~Cheung {\it et al.} [Hadron Spectrum Collaboration],
  JHEP {\bf 1612} (2016) 089
  [arXiv:1610.01073 [hep-lat]].
  
  
\bibitem{Aarts:2014cda}
  G.~Aarts, C.~Allton, T.~Harris, S.~Kim, M.~P.~Lombardo, S.~M.~Ryan and J.~I.~Skullerud,
  JHEP {\bf 1407} (2014) 097
  [arXiv:1402.6210 [hep-lat]].

\bibitem{Aarts:2014nba}
  G.~Aarts, C.~Allton, A.~Amato, P.~Giudice, S.~Hands and J.~I.~Skullerud,
  JHEP {\bf 1502} (2015) 186
  [arXiv:1412.6411 [hep-lat]].


\bibitem{openqcd-fastsum}
J.~Glesaaen and B.~J\"ager,
{\em openQCD-FASTSUM} (v1.0),
\href{http://doi.org/10.5281/zenodo.2216356}{\tt doi.org/10.5281/zenodo.2216356}. 

\bibitem{openqcd}
OpenQCD, \href{http://luscher.web.cern.ch/luscher/openQCD/}{\tt luscher.web.cern.ch/luscher/openQCD/}.

\bibitem{openqcd-hadspec}
  J.~Glesaaen,
  {\em openqcd-hadspec} (v0.1),
  \href{http://doi.org/10.5281/zenodo.2217028}{\tt doi.org/10.5281/zenodo.2217028}.

    


  

\bibitem{Giusti:1998wy}
  L.~Giusti, F.~Rapuano, M.~Talevi and A.~Vladikas,
  Nucl.\ Phys.\ B {\bf 538} (1999) 249
  [hep-lat/9807014].
  
  
  
\bibitem{Aarts:2015mma}
  G.~Aarts, C.~Allton, S.~Hands, B.~J\"ager, C.~Praki and J.~I.~Skullerud,
  Phys.\ Rev.\ D {\bf 92} (2015) no.1,  014503
  [arXiv:1502.03603 [hep-lat]].

\bibitem{Aarts:2017rrl}
  G.~Aarts, C.~Allton, D.~De Boni, S.~Hands, B.~J\"ager, C.~Praki and J.~I.~Skullerud,
  JHEP {\bf 1706} (2017) 034
  [arXiv:1703.09246 [hep-lat]].

\bibitem{Aarts:2018glk}
  G.~Aarts, C.~Allton, D.~De Boni and B.~J\"ager,
  Phys.\ Rev.\ D {\bf 99} (2019) no.7,  074503
  [arXiv:1812.07393 [hep-lat]].
  


\bibitem{Datta:2012fz}
  S.~Datta, S.~Gupta, M.~Padmanath, J.~Maiti and N.~Mathur,
  JHEP {\bf 1302} (2013) 145
  [arXiv:1212.2927 [hep-lat]].


  
  
\end{thebibliography}
\end{document}